\documentstyle[12pt]{article}
\begin{document}
\begin{flushright}
Preprint IHEP 94-92\\
September 1994\\
Submitted to JETP Lett.
\end{flushright}
\begin{center}
{\bf \large $\boldmath{\alpha_S}$ from Spectroscopy of
$\boldmath{\psi}$- and $\boldmath{\Upsilon}$-particles\\
in QCD Sum Rules}
\\
\vspace*{1cm}
V.V.Kiselev\\
{\it Institute for High Energy Physics,\\
Protvino, Moscow Region, 142284, Russia,\\
E-mail: kiselev@mx.ihep.su\\
Fax: +7-095-230-23-37}
\end{center}

\begin{abstract}
In a specific scheme of the QCD sum rules, one gives the estimate of
$\alpha_S\simeq 0.20$ from the data on the masses and leptonic constants of
$\psi$- and $\Upsilon$-particles.
\end{abstract}

The characteristic value of the quark-gluon coupling in QCD for the systems
of the charmonium $(\bar c c)$ and bottomonium $(\bar b b)$ is generally
estimated from the branching fractions of the radiative and leptonic
decays of vector states $Br(Q\bar Q \to \gamma X)\simeq \Gamma(Q\bar Q\to
\gamma g g)/\Gamma(Q\bar Q\to ggg)$, $Br(Q\bar Q \to l^+l^-)\simeq
\Gamma(Q\bar Q\to l^+l^-)/\Gamma(Q\bar Q\to ggg)$ [1], so these ways do not
depend on a modelling of the heavy quarkonium wave function. However, such
estimates contain the uncertainty, related with a model for the gluon
hadronization. For instance, one can think, that the gluons have a nonzero
virtuality of the order of the confinement scale or higher corrections
generate a nonzero effective dynamical mass of the gluon [2]. The
$\alpha_S$ estimates from the total cross sections for the hadronic
production of the $c$- and $b$-quarks also contain the uncertainties, related
with an account of higher corrections in the perturbation theory of QCD
(the $K$-factor), a choice of the quark mass values  and a model-dependence
of the parton distributions. Therefore, one generally supposes
$\alpha_S(\psi,\; \Upsilon)\simeq 0.3\div 0.2$. It would be useful one
to obtain the characteristic $\alpha_S$ value for the $\psi$- and
$\Upsilon$-particles, using another way of the estimate.

To get the $\alpha_S$ estimate, in the present paper we use the relations,
obtained in the framework of the QCD sum rules [3] for the leptonic
constants and masses of the heavy quarkonia.

In the recently offered scheme of the QCD sum rules [4], in the leading
approximation over the inverse heavy quark mass and with an account of the
coulomb-like $\alpha_S/v$-corrections, one has got the relations for the
leptonic constants $f_n$ of the heavy quarkonium $nS$-levels, lying below
the threshold of the decay into the heavy meson pair, [4]
\begin{equation}
\frac{f_n^2}{M_n} = \frac{\alpha_S}{\pi} \; \frac{dM_n}{dn}\;,
\end{equation}
and for the difference of the level masses [5]
\begin{equation}
{M_n-M_1} = \frac{dM_n}{dn}(n=1)\; {\ln {n}}\;.
\end{equation}
Since the heavy quark potential is close to the logarithmic one [6],
the quarkonium level density $dn/dM_n$ does not depend on the heavy
quark flavours. Therefore, with the accuracy up to logarithmic
loop-corrections, one gets, that
\begin{equation}
\frac{f^2}{M} = const.\;,
\end{equation}
and
\begin{equation}
\frac{M_n-M_1}{\ln{n}}= const.
\end{equation}
The data on the $\psi$- and $\Upsilon$-particle spectroscopy show, that
relations (3) and (4) are valid with a good accuracy ($\le 10$\%) [4,5].
Then one can obtain the reliable estimate for the $\alpha_S$ value
\begin{equation}
\alpha_S = \pi\; \frac{f_{1S}^2}{M(1S)}\; \frac{\ln{2}}{M(2S)-M(1S)}\;,
\end{equation}
so that the quantities in the right hand side of eq.(5) are well known
experimentally. From eq.(5) it follows, that
\begin{equation}
\alpha_S(\psi,\; \Upsilon) \simeq 0.20\pm 0.02\;,
\end{equation}
where the error corresponds to the accuracy of the approach as the whole.

If one uses the one-loop expression for the "running" constant
$\alpha_S(\mu) = 2\pi/(11-2n_f/3)\ln{\mu/\Lambda}$, where $n_f=3$ is
the number of light quarks, $\Lambda$ is the one-loop renormalization
invariant, then, accepting the mean value in eq.(6) and
$\mu =(m_\psi+m_\Upsilon)/2$, we get $\Lambda\simeq 0.14$ GeV and
$\alpha_S(\psi)= 0.25\pm 0.03$, $\alpha_S(\Upsilon)= 0.18\pm 0.02$, that is
in a good agreement with estimate (6).

Thus, in the framework of the QCD sum rules and on the basis of the
spectroscopic data, one gets the reliable estimate of $\alpha_S$ for the
$\psi$- and $\Upsilon$-systems.
\vspace*{0.5cm}

\small
\centerline{{\bf References}}
\begin{enumerate}
\item
T.Appelquist, H.D.Politzer, Phys.Rev.Lett. {\bf 34}, 43 (1975);\\
A.DeRujula, S.L.Glashow, Phys.Rev.Lett. {\bf 34}, 46 (1975);\\
V.A.Novikov et al., Phys.Rep. {\bf 41C}, 1 (1978).
\item
M.Consoli, J.H.Field, Phys.Rev. {\bf D49}, 1293 (1994);\\
H.C.Chiang, J.H\" ufner, H.J.Pirner, Phys.Lett. {\bf B324}, 482 (1994).
\item
M.A.Shifman, A.I.Vainstein, V.I.Zakharov, Nucl.Phys. {\bf B147} 385,
448 (1979);\\
L.J.Reinders, H.Rubinstein, T.Yazaki, Phys.Rep. {\bf 127} 1 (1985).
\item
V.V.Kiselev, Nucl.Phys. {\bf B406}, 340 (1993).
\item
V.V.Kiselev, Preprint IHEP 94-74, Protvino (1994).
\item
E.Eichten, Preprint FERMILAB-Conf-85/29-T (1985);\\
C.Quigg and J.L.Rosner, Phys.Lett. {\bf B71}, 153 (1977).
\end{enumerate}
\end{document}